# Finite Temperature QCD with Wilson Quarks: A Study with a Renormalization Group Improved Gauge Action*

Y. Iwasaki[a,b], K. Kanaya[a,b], S. Kaya[b], S. Sakai[c] and T. Yoshié[a,b]

[a]Center for Computational Physics, University of Tsukuba, Ibaraki 305, Japan

[b]Institute of Physics, University of Tsukuba, Ibaraki 305, Japan

[c]Faculty of Education, Yamagata University, Yamagata 990, Japan

Finite temperature transition in lattice QCD with degenerate Wilson quarks is investigated on an $N_t = 4$ lattice, using a renormalization group improved gauge action. We find the following for the $N_F = 2$ case: 1) The transition is smooth for a wide range of the quark mass. 2) The chiral transition is continuous. 3) The chiral condensation well satisfies a scaling relation with the critical exponents of the 3 dimensional $O(4)$ spin model. For $N_F = 3$, we find that the chiral transition is of first order.

## 1. Introduction

In our previous work [1] we investigated finite temperature lattice QCD with two degenerate Wilson quarks using a renormalization group improved gauge action [2] and the standard Wilson quark action. We showed that the unexpected sharp transition observed with the standard gauge action in the intermediate mass range [3] disappears on an $N_t = 4$ lattice with the improved action in the quark mass region where the sharp transition was observed. In this article we extend this work to a wider range of $\beta$ and extensively investigate the dependence of various physical quantities on $\beta$ and quark mass. Among others we check a scaling relation suggested by a universality argument [4]. We also investigate the three flavor case with the improved action.

Numerical simulations for $N_F = 2$ are done on $8^3 \times 4$ and $8^4$ lattices. The critical hopping parameter $K_c$ is determined by $m_\pi^2 = 0$ on the $8^4$ lattice and the lattice spacing $a$ from $m_\rho$ at $K_c$. For $N_F = 3$ we simulate $8^3 \times 4$ and $12^3 \times 4$ lattices.

## 2. $N_F$=2

In Fig. 1-a we show the Polyakov loop obtained with our improved action for $N_t = 4$ as a function of $1/K - 1/K_c$ ($K_c$ is interpolated when data is not available). The change of the Polyakov loop at the finite temperature transition/crossover $K_t$ is very smooth for a wide range of $\beta$. As we decrease $\beta$, $K_t$ moves toward $K_c$ monotonously and the change of the Polyakov loop becomes steeper. These properties are in sharp contrast with those observed for the standard action (Fig. 1-b).

Fig. 2-a shows the pion screening mass squared $m_\pi^2$. As the quark mass decreases, $m_\pi^2$ deviates gradually from that in the low temperature phase. Cusps observed with the standard action[3] (Fig. 2-b) disappear for the whole range of $\beta$. We also observe that the envelope of $m_\pi^2$ on the $N_t = 4$ lattice agrees excellently with $m_\pi^2$ for $N_t = 8$.

It is known that the quark screening mass $m_q$ defined by an axial-vector Ward identity [5,6] exhibits peculiar behavior with the standard action in the high temperature phase at $\beta \lesssim 5.3$: $m_q$ for $N_t = 4$ shows a sharp bend at $K_t$ and the value in the high temperature phase does not agree with that in the low temperature phase[3,1]. As shown in Fig. 3, $m_q$ with the improved action shows no such strange behavior: $m_q$ for $N_t = 4$ changes smoothly from the low temperature phase to the high temperature phase and agrees well with that obtained on the $N_t = 8$ lattice.

It was pointed out[7] that the unexpected sharp

*Talk presented by T. Yoshié at *Lattice '95*.

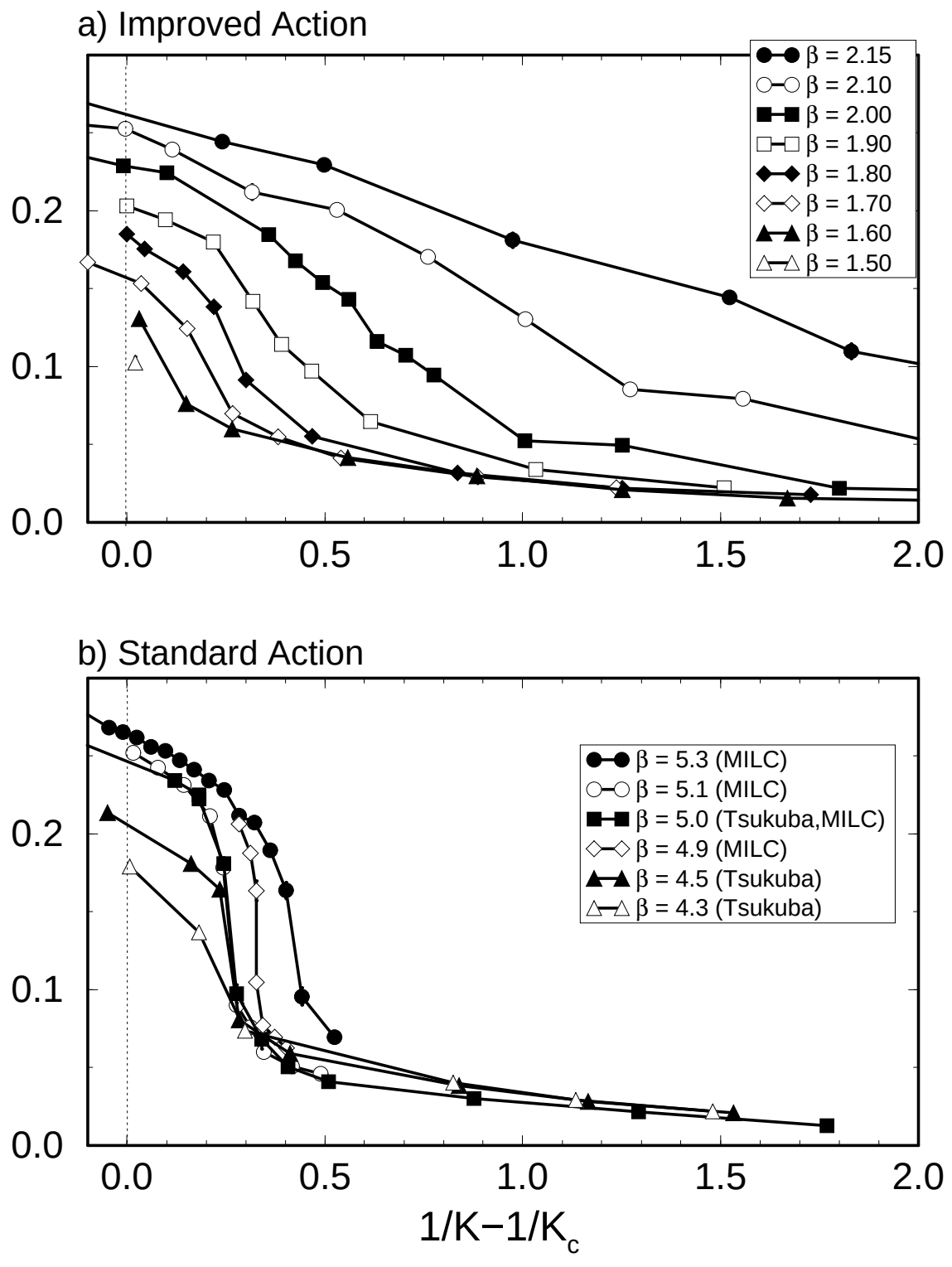

Figure 1. Polyakov loop for $N_t = 4$ versus $1/K - 1/K_c$ obtained with (a) the improved action and (b) the standard action. $\beta = 2.1$ (1.6) for the improved action corresponds in terms of $a^{-1}$ to $\beta = 5.1$ (4.5) for the standard action ($a^{-1} \sim 1.2$ (0.83) GeV).

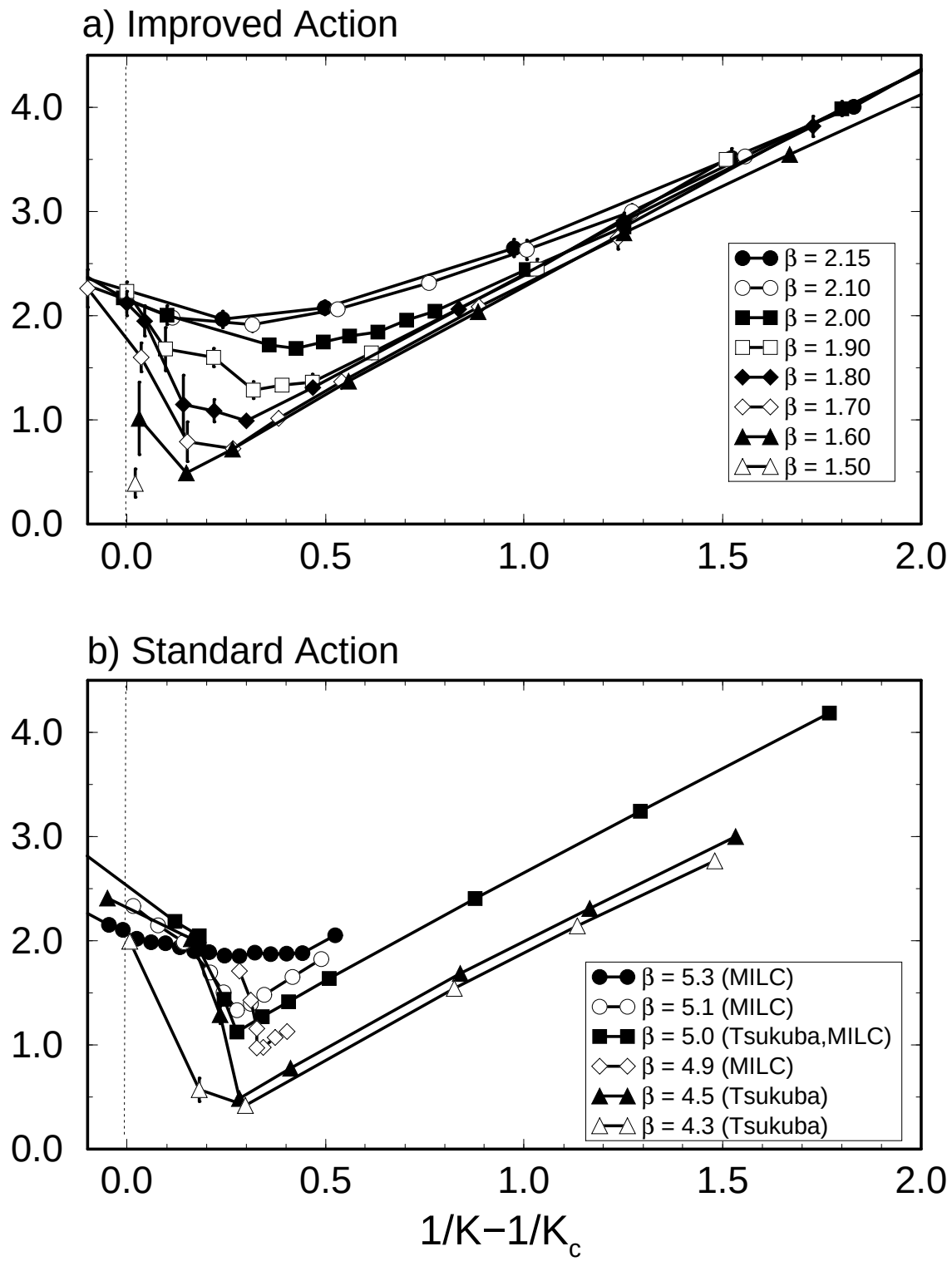

Figure 2. The same as Fig. 1 for $m_\pi^2$.

transition in the intermediate quark mass region with the standard action is probably due to the unusual relation between the $K_c$ line and the $K_t$ line: the $K_t$ line which deviates from the $K_c$ line at the chiral transition point approaches toward the $K_c$ line again at $\beta \sim 5.0$ where the sharp transition was observed. In the case of the improved action, as shown in Fig. 4, the distance between $K_c$ and $K_t$ decreases monotonously toward the crossing point, i.e. the chiral transition point $\beta_{ct}$. In accord with this, the sharp transition disappears with the improved action, as shown above. Thus we conclude that the unexpected sharp transition observed with the standard action is a lattice artifact.

Applying the method described in Ref. [7,8], we have performed simulations on the $K_c$ line in order to locate $\beta_{ct}$ and to determine the order of the transition there. In fig.5 we show $m_\pi^2$ on the $K_c$ line together with that on the $K_t$ line as functions of $\beta$. The $m_\pi^2$ decreases toward zero as $\beta$ decreases and no two state signals are observed. This implies that the chiral transition is continuous. A linear extrapolation of $m_\pi^2$ gives $\beta_{ct} \sim 1.4$. The smoothness of physical quantities on $K_t$ at intermediate masses shown in Figs. 1-a and 2-a strongly suggests that the transition is a crossover there. This is in accordance with the theoretical expectation that a continuous chiral transition turns into a crossover at finite $m_q$.

A universality argument suggests that, if the chiral transition in $N_F = 2$ QCD is of second order, then it belongs to the same universality class as the 3 dimensional $O(4)$ Heisenberg model [4,9]. We can use the resulting scaling relations to make clear the nature of the chiral transition. Smooth behavior of physical quantities with the improved action, in sharp contrast with the case of the standard action, encourages us to test the scaling with Wilson quarks. Taking correspondences of $\beta-\beta_{ct}$, $2m_q$ and the chiral condensation $\langle\bar{\psi}\psi\rangle$ in QCD to the reduced temperature $t$, the magnetic field $h$ and the magnetization $M$, we test the scaling relation [4]

$$M(t,h) = h^{1/\delta} f(th^{-1/\beta\delta}),$$

where $f$ is a scaling function and the $\beta$ and the $\delta$ are critical exponents. (This scaling relation has been tested for staggered quarks using different correspondences of scaling variables [10].)

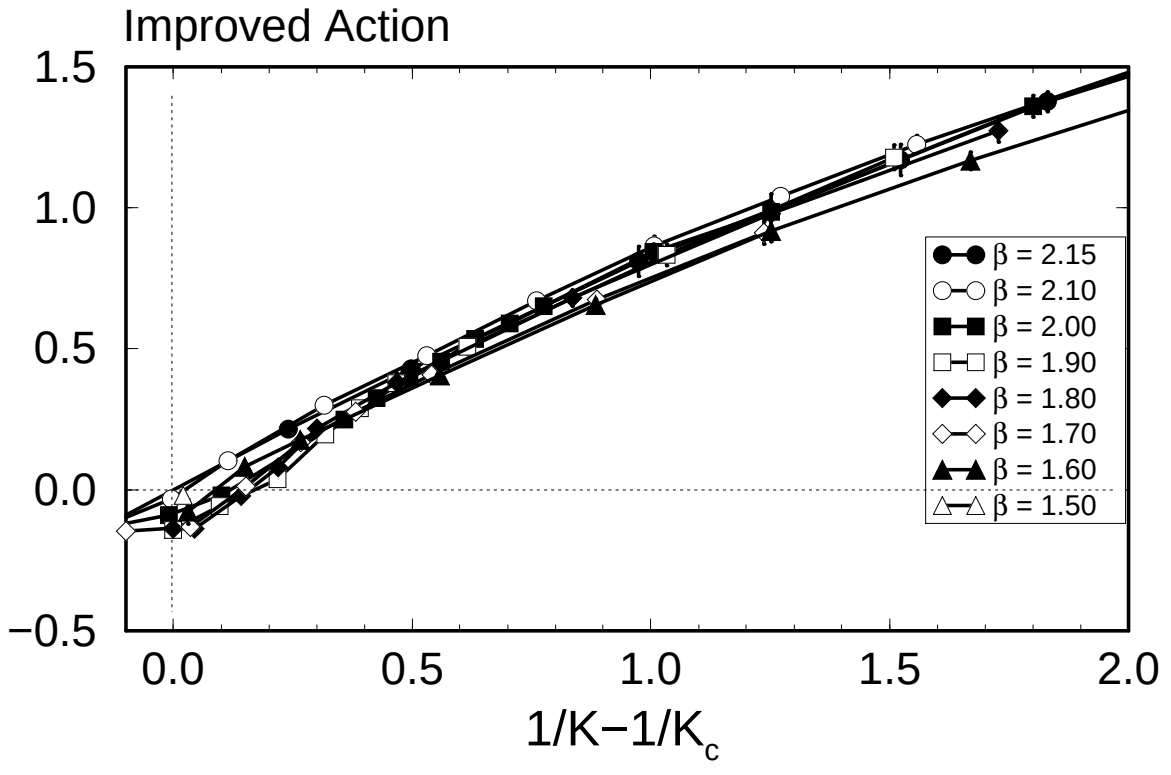

Figure 3. The same as Fig. 1-a for $2m_q$.

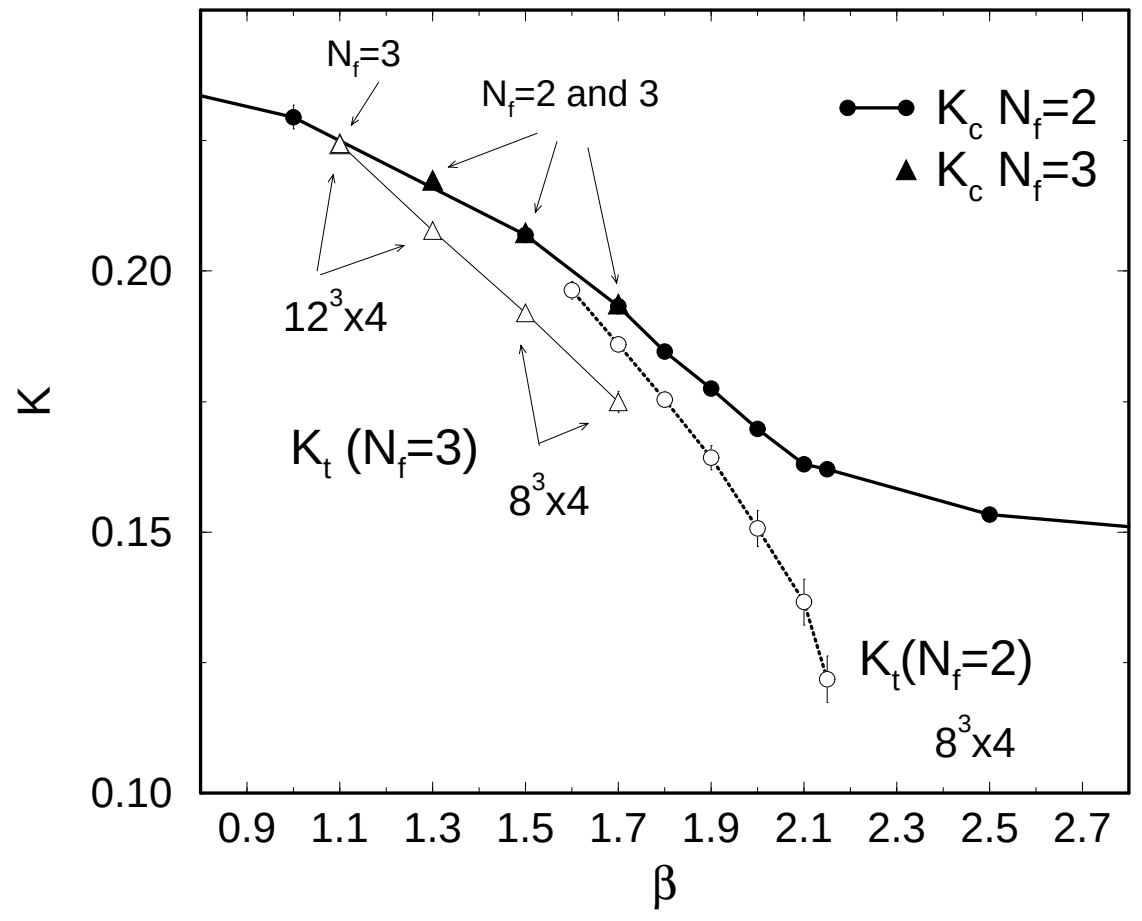

Figure 4. Phase diagram for $N_F = 2$ and $N_F = 3$ with the improved action.

The explicit chiral violation of Wilson fermions leads us to use a subtracted chiral condensation $\langle\bar{\psi}\psi\rangle$ defined by a chiral Ward identity[5]:

$$\langle\bar{\psi}\psi\rangle = 2m_q \sum_x \langle P_5(x) P_5(0)\rangle,$$

where $P_5(x)$ is the pseudo scalar density. (We have dropped the overall renormalization coefficient which plays no role in our study of scaling properties.) We vary $\beta_{ct}$ as a parameter to find a value where the scaling relation is best satisfied with the $O(4)$ critical exponents [11]. Using all data in the range of $\beta = 1.6 - 2.0$ and $0 < 2m_q < 0.9$, we find that data are nicely on a universal curve when we set $\beta_{ct} = 1.34(3)$. Fig. 6 shows the scaling function $Mh^{-1/\delta}$ as a function of the rescaled temperature $th^{-1/\beta\delta}$. We note that this value of $\beta_{ct}$ is consistent with our results for $m_\pi^2$ shown in Fig. 5, because the universality predicts that $m_\pi^2$ on the both lines should bend upward when $\beta$ approaches toward $\beta_{ct}$ [4]. On the other hand, when we fix the exponents to mean field values, the data for $\langle\bar{\psi}\psi\rangle$ are not on a universal curve so well as in the case of $O(4)$ exponents. Furthermore the best fit value $\beta_{ct} = 1.48(3)$ in this case is apparently too large when we consider the results in Fig. 5. Therefore we conclude that our data are consistent with the picture that the chiral transition belongs to the universality class of the 3-dimensional $O(4)$ Heisenberg model. The success of our scaling test strongly suggests that the chiral transition is of second order for $N_F = 2$ in the continuum limit. The direct extraction of each critical exponent and the comparison with that of $O(4)$ model are necessary to make this conclusion more definite.

## 3. $N_F$=3

The phase diagram obtained so far has been given in Fig. 4. Simulations are made on the $K_c$

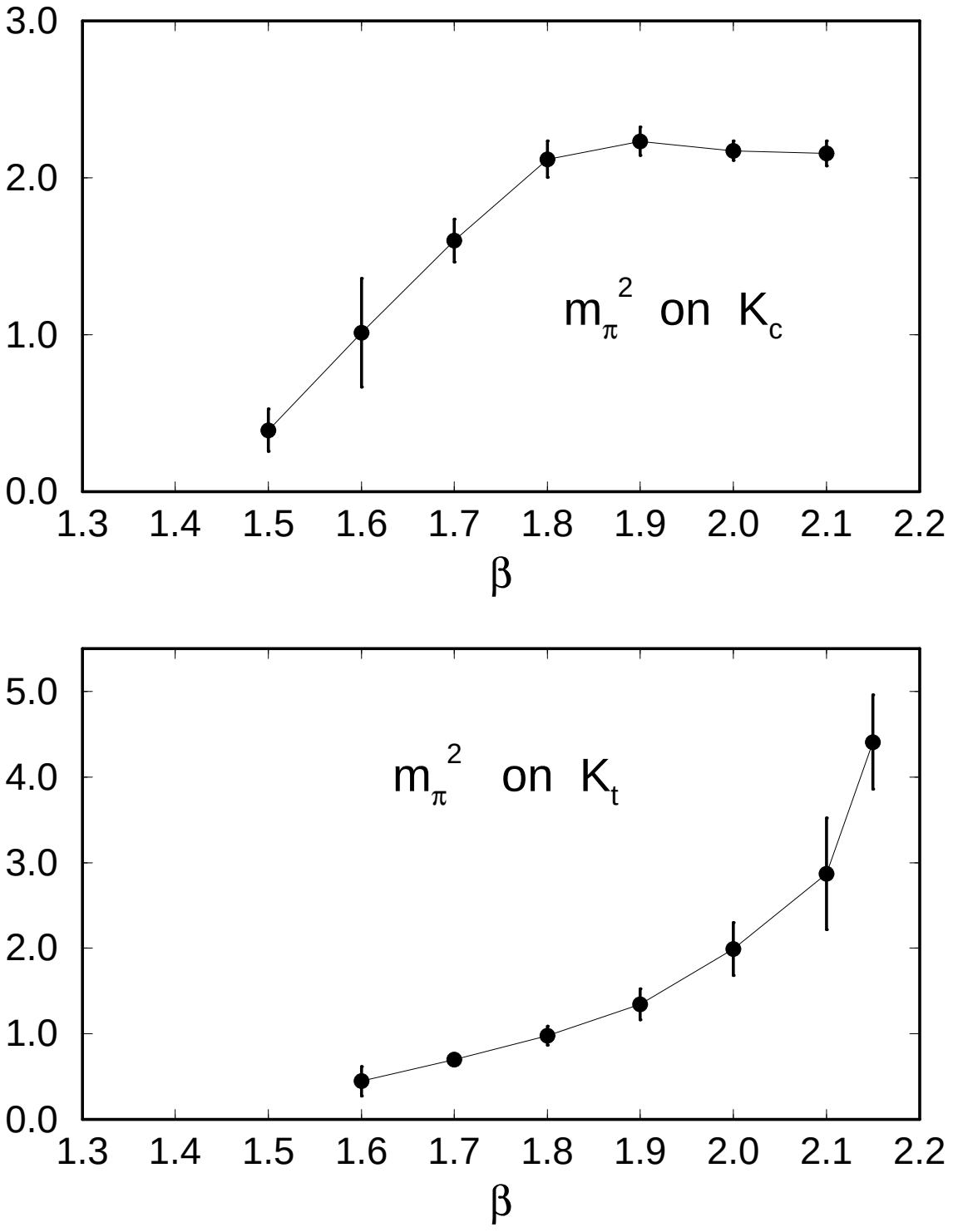

Figure 5. $m_\pi^2$ on the $K_c$ line and that on the $K_t$ line versus $\beta$ for $N_F = 2$.

line at $\beta = 1.1, 1.2$ and $1.3$ on a $12^3 \times 4$ lattice. Using the method described in Ref. [7,8], we find $\beta_{ct} \sim 1.1$ and observe two-state signals at $\beta = 1.1$. Therefore we conclude that the chiral transition is of first order. Simulations for finite quark mass are in progress.

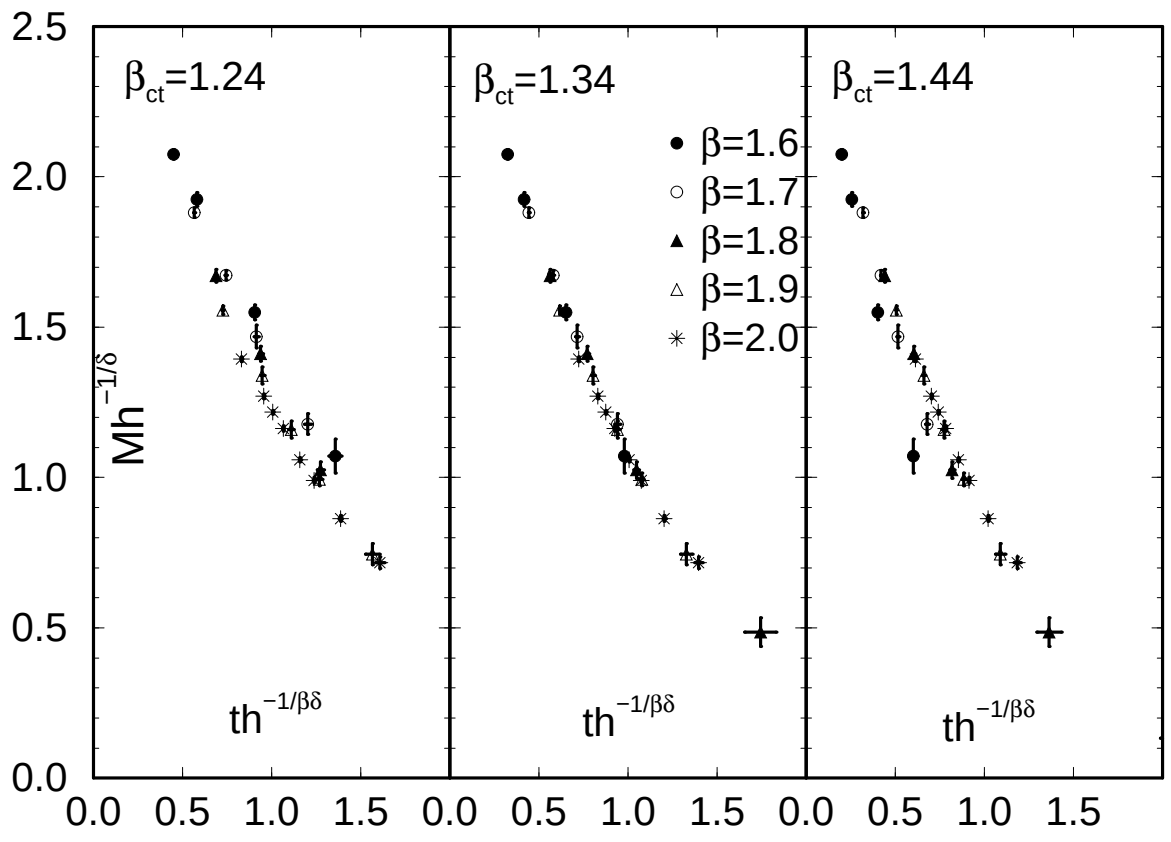

Figure 6. A test of a scaling relation for the chiral condensation. See text for details.

## Acknowledgements

The simulations are performed with Fujitsu VPP500/30 at the University of Tsukuba. This work is supported in part by the Grant-in-Aid of Ministry of Education, Science and Culture (No. 07NP0401, 07640375 and 07640376).